\documentclass[preprintnumbers,amsmath,amssymb,preprint]{revtex4}

\usepackage{amsmath}
\usepackage{graphicx}% Include figure files
\usepackage{dcolumn}% Align table columns on decimal point
\usepackage{bm}% bold math
\usepackage[usenames]{color}
\usepackage{multirow}

\begin{document}

\title{SOME FEATURES OF THE STATE-SPACE TRAJECTORIES FOLLOWED BY ROBUST ENTANGLED
FOUR-QUBIT STATES DURING DECOHERENCE}

\author{A.P. Majtey$^{1}$,
A. Borras$^{1,\,2}$, A.R. Plastino$^{3,\,4}$, M. Casas$^{1}$, and
A. Plastino$^{4}$} \affiliation{ $^1$Departament de F\'{\i}sica
and IFISC, Universitat de
les Illes Balears, 07122 Palma de Mallorca, Spain. \\
$^2$Kavli Institute of Nanoscience, Delft University of Technology,\\
Lorentzweg 1, 2628 CJ Delft, The Netherlands. $^3$Instituto Carlos
I de F\'{\i}sica Te\'orica y Computacional,
Universidad de Granada, Granada, Spain. \\\\
$^4$National University La Plata UNLP-Conicet, C.C. 727, 1900 La
Plata, Argentina.}

\begin{abstract}
In a recent work (Borras {\it et al.}, Phys. Rev. A {\bf 79},
022108 (2009)), we have determined, for various decoherence
channels, four-qubit initial states exhibiting the most robust
possible entanglement. Here we explore some geometrical features
of the trajectories in state space generated by the decoherence
process, connecting the initially robust pure state with the
completely decohered mixed state obtained at the end of the
evolution. We characterize these trajectories by recourse to the
distance between the concomitant time dependent mixed state and
different reference states.
\end{abstract}

\maketitle

\section{Introduction}
Entanglement and decoherence are two closely related phenomena
that lie at the core of quantum physics \cite{BZ06,NC00,S05}.
Entanglement is nowadays regarded as the most distinctive feature
of quantum mechanics, and in recent years it has been the subject
of intense and increasing research efforts. The phenomenon of
decoherence consists, basically, of a set of effects arising from
the interaction (and concomitant entanglement development) between
quantum systems and their environments \cite{NC00,S05}. Every
physical system is immersed in an environment and interacts with
it in some way. The effect due to this interaction is one of the
main obstacles to the practical implementation of quantum
technologies based on the controlled manipulation of entangled
states such as quantum computing \cite{NC00}. The decoherence
process leads the system from a pure state to a (usually less
entangled) mixed state. This decay of entanglement has recently
attracted the interest of many researchers
\cite{BMPCP08,SK02,CMB04,DB04,HDB05,ACCAD08,AlmeidaEtAl07,SallesEtAl08,AolitaEtAl09,GBB08,BGB08,CTP09}.
It has also been shown that in some cases entanglement can vanish
in finite times. This phenomenon is known as \emph{entanglement
sudden death} and it has been the focus of numerous contributions
\cite{YE04,YE09,QJ08,Weinstein09,SB09,ZX09}. This abrupt
disappearance of entanglement is closely related to the
\emph{sudden birth of entanglement} between the reservoirs
\cite{LRLSR08}.

Recently, several dynamical properties of entanglement, like
asymptotic birth of entanglement and entanglement sudden death,
have been discussed from a geometrical point of view
\cite{TC07,DTC08}. These geometrical interpretations allow for the
explanation of the necessary and sufficient conditions for the
last phenomenon to take place \cite{TC07}. Various examples
according to different possibilities for the geometrical details
of the set of time asymptotic states are provided in \cite{DTC08}.

We have recently studied the decay of entanglement under different
paradigmatic noisy channels, identifying the initial states
exhibiting the most robust entanglement \cite{BMPCP08}.

The aim of the present contribution is to explore characteristic
traits of those state space trajectories associated to robust
states, with emphasis on the study, from a geometrical point of
view, of optimal-ones. This will hopefully shed some light on the
existence and behaviour of  robust states. The paper is organized
as follows. In Section II we briefly review the local decoherence
models for multi-qubit systems and the entanglement and distance
measures that will be used in the present work. In Section III we
investigate the evolution trajectories in state space of highly
entangled four-qubit states. Finally, some conclusions are drawn
in Section IV.

\section{Preliminaries}

The systems under consideration in the present study consist of an
array of $N$ independent qubits initially entangled due to a
previous, arbitrary interacting process. Each qubit in the
composite system is coupled to its own environment. In such a
local-environment formulation there is no communication among
qubits and the entanglement between the subsystems cannot increase
because of the locality of the involved operations. We assume that
all qubits are affected by an identical decoherence process. The
dynamics of any of these qubits is governed by a master equation
from which one can obtain  a completely positive trace-preserving
map $\varepsilon$ which describes the corresponding evolution:
$\rho_i (t) = \varepsilon \rho_i (0)$. In the Born-Markov
approximation these maps (or channels) can be described using its
Kraus representation

\begin{equation}
 \varepsilon_i \  \rho_i(0) = \sum_{j=1}^{M} E_{j \, i} \
 \rho_i (0) \  E_{j \, i}^\dag,
\end{equation}

\noindent where $E_j\,\,\,j=1,\ldots, M$ are the so called Kraus
operators, $M$ being the number of operators needed to completely
characterize the channel \cite{K83}. Using the Kraus operators
formalism it is possible to describe the evolution of the entire
N-qubit system,

\begin{equation}
 \rho(t)= \varepsilon \ \rho(0) \  = \  \sum_{i...j} E_{i \, 1} \otimes \ldots
 \otimes E_{j \, N} \ \rho(0) \  [ E_{i \, 1} \otimes \ldots \otimes E_{j \, N}]^{\dag}.
\end{equation}

\subsection{Decoherence models}
We concentrate our efforts in the study of the following family of
decoherence channels: the bit flip (BF), phase flip (PF), and
bit-phase flip (BPF). These channels represent all the possible
errors in quantum computation, the usual bit flip
$0\leftrightarrow 1$, the phase flip, and the combination of both,
bit-phase flip. The corresponding pair of Kraus operators $
\{E_0,E_1 \} $ for each channel is given by:

\begin{equation}\label{BF-kraus}
E_0 = \sqrt{1-p/2}I, E_1^i=\sqrt{p/2}\sigma_i;
\end{equation}

\noindent where $I$ is the identity matrix, $\sigma_i$ are the
Pauli matrices and $i=x$ give us the bit flip, $i=z$ the phase
flip, and $i=y$ the bit-phase flip. Following Salles {\it et al.}
\cite{SallesEtAl08}, the factor 2 in Eq. (\ref{BF-kraus})
guarantees that at $p=1$ the ignorance about the occurrence of an
error is maximal, and as a consequence, the information about the
state is minimum. The results obtained with the phase damping
channel are the same than those of the phase flip channel,
actually they can be shown to represent the same process, under a
proper change of variables \cite{NC00}.

We also consider the depolarizing channel (D) which can be viewed
as a process in which the initial state is mixed with a source of
white noise with probability $p$. For a $d$-dimensional quantum
system, it can be expressed as

\begin{equation}\label{D-map}
\varepsilon \, \rho \, = \, \frac{p}{d} \, I + (1-p) \, \rho
\end{equation}

\noindent The Kraus operators for this process, including all
Pauli matrices are

\begin{equation}\label{D-kraus}
E_0 \, = \, \sqrt{1 - p'} I \, ; \ \ \ E_i \, = \,
\sqrt{\frac{p'}{3}} \sigma_i
\end{equation}

\noindent where $p' = \frac{3p}{2}$. Under this process the state
turns separable after a finite time, being this channel the only
one considered in this work exhibiting the phenomenon of
entanglement sudden death. We remember here that due to the high
symmetry of the depolarizing channel the evolution of the
entanglement depends only on the amount of entanglement of the
initial state. Then, according to this process, we obtain
equivalent evolutions of the entanglement for initial states which
are equivalent under local unitary operation \cite{BMPCP08}. Some
results for this channel will be commented at the end of Sec. III.

\subsection{Multipartite entanglement quantification}
One of the most popular measures proposed to quantify multipartite
entanglement is based on the use of a bipartite measure, which is
averaged over all possible bipartitions of the system. It is
mathematically expressed in the fashion

\begin{eqnarray}\label{Ent-measure}
E &=& \frac{1}{[N/2]} \sum_{m=1}^{[N/2]} E^{(m)}, \\
E^{(m)} &=& \frac{1}{N_{bipart}^m} \sum_{i=1}^{N_{bipart}^m} E(i).
\label{Entsub}
\end{eqnarray}
\noindent Here, $E(i)$ stands for the entanglement associated with
one single bipartition of the $N$-qubit system. The quantity
$E^{(m)}$ gives the average entanglement between subsets of $m$
qubits and the remaining $N-m$ qubits that constitute the system.
The average is performed over the ensuing $N_{bipart}^{(m)}$
nonequivalent bipartitions. If one uses the linear entropy $S_L$
of the reduced density matrix of the smaller bipartition to
compute $E(i)$, $E_L^{(1)}$ turns out to be the well known
Meyer-Wallach multipartite entanglement measure \cite{MW02}. This
measure was later generalized by Scott to the case where all
possible bipartitions of the system were considered
\cite{Scott04}. The Meyer-Wallach multipartite entanglement
measure has recently been related to the regularized quantum
Fisher information measure which gives the estimation of the
strength of low-noise locally depolarizing channels \cite{BM08}.

We will use the negativity as our bipartite measure of
entanglement because we are dealing with mixed states. The
negativity is proportional to the sum of the negative eigenvalues
$\alpha_i$ of the partial transpose matrix associated with a given
bipartition. The properly normalized negativity reads

\begin{equation}\label{neg}
Neg=\frac{2}{2^m-1} \sum_i |\alpha_i|.
\end{equation}

\subsection{Distance measures}

Distance measures between quantum states constitute important
tools in quantum information theory
\cite{MLP05,LMBCP08,BHT08,VP98,CTMAB08}. In the present work, in
order to characterize the trajectories of decohered states, we
compute the distances between the state of interest and several
reference states, such as the initial, final, and maximally mixed
ones. We measure the distance between two different quantum mixed
states by recourse to the quantum Jensen Shannon divergence
(QJSD), which can be defined in terms of the relative entropy as
\cite{MLP05}
\begin{equation}
d_{JS}(\rho,\sigma) =
\frac{1}{2}\left[S\left(\rho,\frac{\rho+\sigma}{2}\right)+
S\left(\sigma,\frac{\rho+\sigma}{2}\right)\right]
\label{JS-relative-entropy}
\end{equation}
that can be recast in terms of the von Neumann entropy $H_N(\rho)
= - Tr(\rho \log \rho)$ in the fashion
\begin{equation}
d_{JS}(\rho,\sigma) = H_N\left(\frac{\rho +
\sigma}{2}\right)-\frac{1}{2} H_N(\rho) - \frac{1}{2} H_N(\sigma)
\label{JS-vN}
\end{equation}
The metric character of the square root of the QJSD has been
ascertained recently for pure states, and strong numerical
evidences have also been found for the mixed states case
\cite{LMBCP08,BHT08}. To avoid an exclusive dependence on the
above  distance measure we also use the Hilbert-Schmidt distance
\cite{VP98}.

\begin{equation}\label{dHS}
d_{HS}(\rho,\sigma) \, = \ {\|\rho - \sigma \|}_{HS}^2 \, = \,
Tr[(\rho-\sigma)^2].
\end{equation}
\noindent We will be able to appreciate the fact that the results
obtained with both distance measures are qualitatively the same.

\subsection{Robust Maximally Entangled Four-Qubit States}

\begin{figure}
\begin{center}
\vspace{0.5cm}
\includegraphics[scale=1,angle=0]{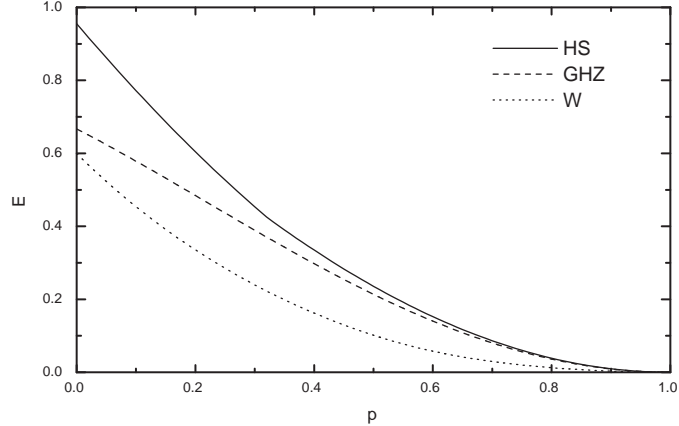}
\caption{Entanglement evolution for several 4 qubits states under
the action of the BF channel. All shown quantities are
dimensionless. \label{4qbrobPF}}
\end{center}
\end{figure}

In a previous work we have performed an iterative numerical search
based on the simulating annealing algorithm and determined  the
initial states that evolve (via the just described noisy channels)
to mixed states with the larger possible amount of entanglement
\cite{BMPCP08}. The proof by Higuchi and Sudbery that a four-qubit
pure state exhibiting the theoretically maximum amount of
entanglement(that is, having all its marginal density matrices
maximally mixed) does not exist constituted a landmark in the
study of multiqubit entanglement \cite{HS00}. In Ref. \cite{HS00},
a promising candidate for  achieving the maximally entangled state
status  was also proposed, namely,

\begin{eqnarray}\label{HS-state}
| \Psi_{rob}^{4} \rangle \, &=& \,|HS\rangle \, = \,
\frac{1}{\sqrt{6}} \Bigl[ |1100\rangle+ |0011\rangle +\\ \nonumber
&\omega& \Bigl(|1001\rangle + |0110\rangle \Bigr) + \omega^2
\Bigl(|1010\rangle + |0101\rangle\Bigr) \Bigr],
\end{eqnarray}

\noindent with $\omega= -\frac{1}{2}+ \imath \frac{\sqrt{3}}{2}$.
Such conjectures have later received support from several
numerical studies \cite{BPBZCP07,BCPP08,FFPP08}. In our previous
work, the $|HS\rangle$ was found to be a robust state. The concept
of robustness used in this work can be easily explained from Fig.
\ref{4qbrobPF}. The $|HS\rangle$ state is considered to be the
most robust state because, for any value of $p$, it leads to
decohered states exhibiting more entanglement than those
associated with other initial states. In this graph we plot the
entanglement-evolution of the $|HS\rangle$ state under the action
of the BF channel and compare it to the entanglement decay of the
well-known entangled states $|GHZ\rangle$ and $|W\rangle$

\begin{eqnarray}\label{GHZ-W}
|GHZ\rangle &=& \frac{1}{\sqrt{2}} (|0000\rangle + |1111\rangle), \\
|W\rangle &=& \frac{1}{2} (|0001\rangle + |0010\rangle +
|0100\rangle + |1000\rangle).
\end{eqnarray}
We also found that, for six qubit systems the robust state
$|\Psi_{rob}^6 \rangle$ turns out to be precisely the maximally
entangled state encountered  by some of the authors of a previous
work \cite{BPBZCP07}. For five qubit systems, the robust state
$|\Psi_{rob}^5 \rangle$ that we found is not as good as the one
detected in the 4 or 6 qubits instance. For BF and BPF channels
its entanglement becomes lower than that of other states for large
$p$ values \cite{BMPCP08}. The entanglement decays of
$|\Psi_{rob}^4 \rangle$ and $|\Psi_{rob}^6 \rangle$ are quite
similar and their entanglements are always larger than that of any
other state tested in our samplings.

\begin{figure}
\begin{center}
\vspace{0.5cm}
\includegraphics[scale=1,angle=0]{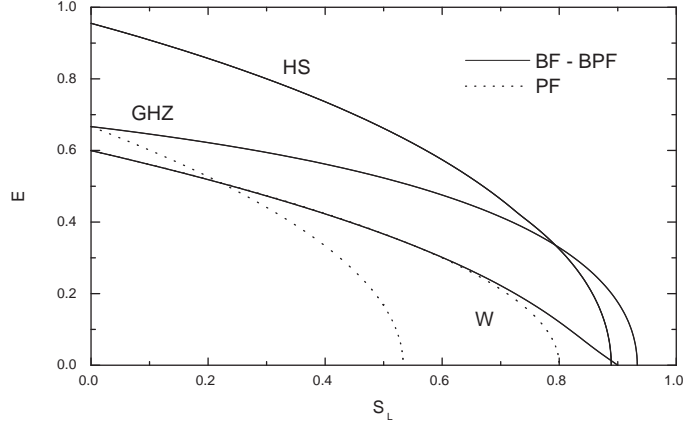}
\caption{Entanglement evolution as a function of the linear
entropy of 4 qubits representative states under phase flip, bit
flip, and bit phase flip decoherence models. All depicted
quantities are dimensionless.\label{EvsSL}}
\end{center}
\end{figure}

\section{Optimal trajectories}
In order to compare the behaviour of the entanglement decay for
the different channels we compute it in terms of the degree of
mixedness, given by the linear entropy of the density matrix of
the (evolved) system

\begin{equation} \label{eq:dec-mix.entlin}
S_L(\rho) \, = \, \frac{N}{N-1} (1-Tr[\rho^2]),
\end{equation}
\noindent where $N=2^n$ and n is the number of qubits. Here, we
display only the results corresponding to four qubit systems,
although similar results are obtained for systems with different
number of qubits.

\begin{figure}
\begin{center}
\vspace{0.5cm}
\includegraphics[scale=1,angle=0]{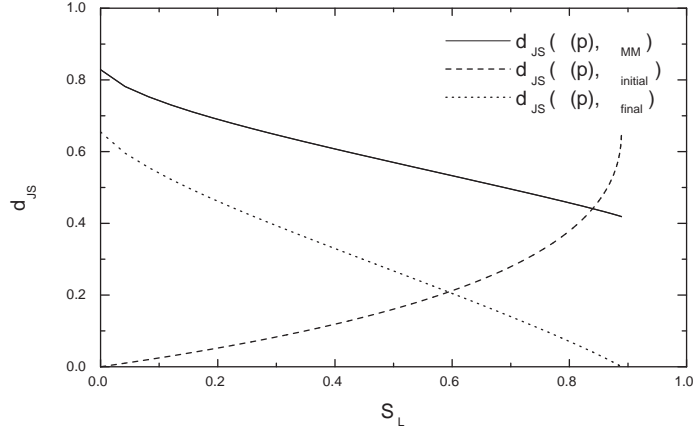}
\caption{QJSD between the decohered state (initially
$|\Psi_{rob}^4 \rangle$) and reference states: maximally mixed
state (solid line), initial robust state (dashed line), and final
separable state (dotted line). All plotted quantities are
dimensionless.\label{QJS-states}}
\end{center}
\end{figure}

We study the decay of entanglement for different initial states
undergoing  several decoherence processes. We note that, for those
channels for which a given state is robust, the decay of
entanglement in terms of the degree of mixedness is the same. In
Fig. \ref{EvsSL} we plot the decay of entanglement for different
four qubits initial states  for BF, PF, and BPF decoherence
channels. The entanglement evolution for the $|HS\rangle$ state
coincides for the three considered channels. Remember that the
$|HS\rangle$ state was found to be robust under the action of the
three channels studied in this work. In contraposition, the
entanglement evolution of the $|GHZ\rangle$ and $|W\rangle$ states
under the PF channel is not equivalent to that under BF and BPF
processes because these states are not robust for the former
development.

We also compute the distances between the states (at any given
time)  and some reference states using the QJSD. We considered as
reference states the initial, final, and the maximally mixed state
(MM). In Fig. \ref{QJS-states} we plot the resultant distances
between i) the mixed state obtained at each time step of the
evolution of the inial robust pure state of  4 qubits and ii) the
reference states for the previously mentioned decoherence
processes. Similar coincident curves (not shown) are obtained for
$|GHZ\rangle$ and $|W\rangle$ states if we consider only the BF
and  BPF channels. According to both graphs the state with robust
entanglement (w.r.t. several decoherence channels) apparently
evolves in the same manner under the action of such channels. It
is important to note that these trajectories are not actually the
same, they are just equivalent.

\begin{figure}
\begin{center}
\vspace{0.5cm}
\includegraphics[scale=0.6,angle=0]{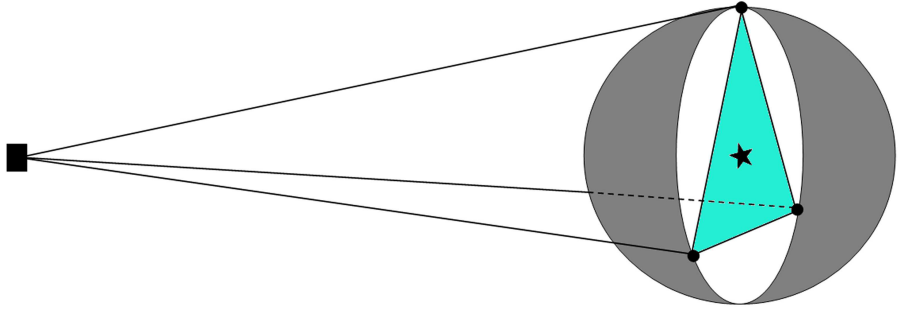}
\caption{Equivalent trajectories in Hilbert space, the initial
robust 4 qubits state is represented by a black square and the
final states corresponding to different decoherence processes are
represented by black spots at the border of the set of separable
states (grey sphere). The star in the center of the sphere
represents the MM state. \label{esquema}}
\end{center}
\end{figure}

Finally, we compute the distances between the three final states.
These resulting states generated by the PF, BF, and BPF channels
when acting upon the initial state $|\Phi_{rob}\rangle$ turn out
to be equidistantly distributed, i.e., the distance between any
pair of them is always the same. The distance from any of them to
the maximally mixed state or to the initial state is also always
the same. The overall picture is displayed in Fig. \ref{esquema}.
The initial state is represented by the black square, and the
three final states are represented by the black spots. These final
states are placed at the border of the set of separable states,
represented by the grey sphere. The star placed in the middle of
the sphere denotes the maximally mixed state. The three decohered
trajectories are different but equivalent, and a nice symmetrical
configuration is observed.

These results can be extended to robust states in higher
dimensions and also to the GHZ and W states under the action of
the BF and BPF channels. The results obtained by using the
Hilbert-Schmidt distance instead of the QJSD are equivalent. The
most representative distances that define the geometry shown in
Fig. 3 are given in the following table:

\vspace{0.5cm}
\begin{center}
\begin{tabular}{|c|c|c|}
\hline
  states & $d_{JS}$ & $d_{HS}$ \\
\hline
  initial-final & 0.6548 & 0.9129 \\
  initial-MM & 0.8285 & 0.9682 \\
  final-MM & 0.4188 & 0.3227 \\
  final-final & 0.6352 & 0.4546 \\\hline
\end{tabular}
\end{center}
\vspace{0.5cm}

\begin{figure}
\begin{center}
\vspace{0.5cm}
\includegraphics[scale=1,angle=0]{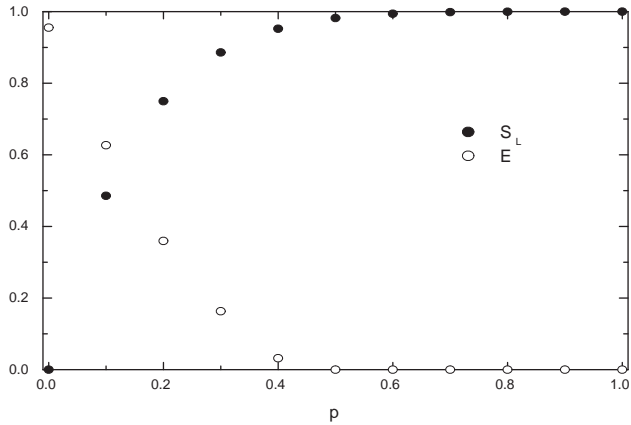}
\caption{Entanglement evolution (empty dots) and linear entropy of
the time evolved state (filled dots) for $10^3$ maximally
entangled states obtained from $|HS\rangle$ by applying local
unitary transformations, under depolarizing channel. All plotted
quantities are dimensionless.\label{E-SL_vs_p}}
\end{center}
\end{figure}

As already mentioned, due to its highly symmetric character, the
depolarizing channel does not have a single robust state. Any
maximally entangled state will be robust for this decoherence
process. All the associated trajectories in state space leading to
these maximally entangled states are equivalent, and the sudden
death of the entanglement always occurs for states  with the same
degree of mixedness. Systems undergoing  this process evolve to a
single asymptotic state, i.e., all initial states evolve
asymptotically to the MM state. Moreover, they do it following
equivalent trajectories. Fig. \ref{E-SL_vs_p} depicts the decay of
the entanglement and the linear entropy of the decohered state for
$10^3$ initial maximally entangled states (equivalent under
unitary operations) due to the action of the depolarizing channel.

\section{Conclusions}

In the present work we have investigated the state space
trajectories  associated to previously determined four-qubit
robust states undergoing different decoherence processes. We have
characterized these trajectories by computing the distance between
the time dependent state resultant from the decoherence process
and some fixed states, namely the initial, final, and maximally
mixed (MM) states. We have shown that, for states that are robust
under decoherence, i.e., those with the maximal amount of
entanglement during the evolution, the trajectories' aspect  may
look different. However, by reference to Fig. 4, one detects a
significant degree of equivalence and symmetry among them.

In the case of depolarizing decoherence processes the final state
is the MM-one, but the system becomes separable before reaching
it, i.e., a sudden death of entanglement takes place for this
process. According to the symmetry of this channel, all maximally
entangled states are robust, and they evolve in equivalent
fashion. Because of this equivalence, the entanglement's sudden
death for any initial maximally entangled state is always
characterized by the same degree of mixedness.

\section*{Acknowledgements} This work was partially supported by the
MEC grant FIS2008-00781 (Spain), by the Project FQM-2445 of the
Junta de Andalucia (Spain), by FEDER (EU), and by Conicet
(Argentine Agency). AB acknowledges support from MEC through FPU
fellowship AP-2004-2962 and APM acknowledges support of MEC
contract SB-2006-0165.


\begin{thebibliography}{}

\bibitem{BZ06} I. Bengtsson  and K. Zyczkowski,
{\it Geometry of Quantum States: An Introduction to Quantum
Entanglement} (Cambridge University Press, Cambridge, 2006)

\bibitem{NC00} M. A. Nielsen and I. L. Chuang,
{\it Quantum Computation and Quantum Information} (Cambridge
University Press, Cambridge, 2000).

\bibitem{S05} M. Schlosshauer, {\it Rev. Mod. Phys.} {\bf 76} (2005) 1267.

\bibitem{BMPCP08} A. Borras {\it et al.}, {\it Phys. Rev. A} {\bf 79} (2009)
022108.

\bibitem{SK02} C. Simon and J. Kempe, {\it Phys. Rev. A} {\bf 65} (2002) 052327.

\bibitem{CMB04} A.R.R. Carvalho {\it et al.}, {\it Phys. Rev. Lett.} {\bf 93} (2004) 230501.

\bibitem{DB04} W. D\"{u}r and H.J. Briegel, {\it Phys. Rev. Lett.} {\bf 92} (2004) 180403.

\bibitem{HDB05} M. Hein {\it et al.}, {\it Phys. Rev. A} {\bf 71} (2005) 032350.

\bibitem{ACCAD08} L. Aolita {\it et al.}, {\it Phys. Rev. Lett.} {\bf 100}
(2008) 080501.

\bibitem{AlmeidaEtAl07} M.P. Almeida {\it et al.}, {\it Science} {\bf 316} (2007) 579.

\bibitem{SallesEtAl08} A. Salles {\it et al.}, {\it Phys. Rev. A} {\bf 78} (2008) 022322.

\bibitem{AolitaEtAl09} L. Aolita {\it et al.}, {\it Phys. Rev. A} {\bf 79} (2009) 032322.

\bibitem{GBB08} O. G\"{u}hne {\it et al.}, {\it Phys. Rev. A} {\bf 78} (2008) 060301.

\bibitem{BGB08} F. Bodoky {\it et al.}, {\it J. Phys.: Cond. Mat.} {\bf 21} (2009) 395602.

\bibitem{CTP09} S. Campbell {\it et al.} {\it New J. Phys.} {\bf 11} (2009)
073039.

\bibitem{YE04} T. Yu and J.H. Eberly, {\it Phys. Rev. Lett.} {\bf 93}(2004) 140404.

\bibitem{YE09} T. Yu and J.H. Eberly, {\it Science} {\bf 323} (2009) 598.

\bibitem{QJ08} A. Al-Qasimi and D. F. V. James, {\it Phys. Rev. A} {\bf 77} (2008) 012117.

\bibitem{Weinstein09} Y.S. Weinstein, {\it Phys. Rev. A} {\bf 79} (2009) 012318.

\bibitem{SB09} I. Sainz and G. Bjork, {\it Int. J. Quant. Inf.} {\bf 7}
(2009) 245.

\bibitem{ZX09} Y-J. Zhang and Y-Y. Xia, {\it Int. J. Quant. Inf.} {\bf
7} (2009) 949.

\bibitem{LRLSR08} C.E. L\'{o}pez {\it et al.}, {\it Phys. Rev. Lett.} {\bf 101} (2008) 080503.

\bibitem{TC07} M. O Terra Cunha, {\it New J. Phys} {\bf 9} (2007) 237.

\bibitem{DTC08} R.C. Drumond and M.O. Terra Cunha, {\it J. Phys. A: Math. Theor.} {\bf 42} (2009) 285308.

\bibitem{K83} K. Kraus, {\it States, Effect, and operation: Fundamental Notions in Quantum
Theroy} (Springer-Verlag, Berlin, 1983).

\bibitem{MW02} D.A. Meyer and N.R. Wallach, {\it J. Math. Phys.} {\bf 43} (2002) 4273.

\bibitem{Scott04} A.J. Scott, {\it Phys. Rev. A} {\bf 69} (2004) 052330.

\bibitem{BM08} S. Boixo and A. Monras, {\it Phys. Rev. Lett.} {\bf 100} (2008) 100503.

\bibitem{MLP05} A.P. Majtey {\it et al.}, {\it Phys. Rev. A} {\bf 72} (2005) 052310.

\bibitem{LMBCP08} P.W. Lamberti {\it et al.}, {\it Phys. Rev. A} {\bf 77} (2008) 052311.

\bibitem{BHT08} J. Bri\"{e}t and P. Harremo\"{e}s, {\it Phys. Rev. A} {\bf 79} (2009) 052311.

\bibitem{VP98}  V. Vedral and M. Plenio, {\it Phys. Rev. A} {\bf 57} (1998) 1619.

\bibitem{CTMAB08} J. Calsamiglia {\it et al.}, {\it Phys. Rev. A} {\bf 77} (2008) 032311.

\bibitem{HS00} A. Higuchi and A. Sudbery, {\it Phys. Lett. A} {\bf 273} (2000) 213.

\bibitem{BPBZCP07} A. Borras {\it et al.}, {\it J. Phys. A: Math. Gen.} {\bf 40}
(2007) 13407.

\bibitem{BCPP08} A. Borras {\it et al.}, {\it Int. J. Quant. Inf.} {\bf 6} (2008) 605.

\bibitem{FFPP08} P. Facchi {\it et al.}, {\it  Phys. Rev. A} {\bf 77} (2008) 060304(R).

\end{thebibliography}
\end{document}